\documentclass[aps,pre,twocolumn,galleyproof]{revtex4}

\usepackage{rotating}
\usepackage{graphicx}
\usepackage{amssymb,amsfonts,amsmath}
\usepackage{dcolumn}
\usepackage{natbib}
\usepackage[all]{xy}
\usepackage{booktabs}
\usepackage{multirow}
\usepackage{color}
\usepackage{amsmath,epsfig}
\usepackage{url}

\begin{document}

\title{Bank-firm credit network in Japan. An analysis of a bipartite network}

\author{Luca Marotta}
\affiliation{Dipartimento di Fisica e Chimica, Universit\`a di Palermo, Viale delle Scienze, Ed. 18, I-90128, Palermo, Italy}

\author{Salvatore Miccich\`e}
\affiliation{Dipartimento di Fisica e Chimica, Universit\`a di Palermo, Viale delle Scienze, Ed. 18, I-90128, Palermo, Italy}

\author{Yoshi Fujiwara}
\affiliation{Graduate School of Simulation Studies, The University of Hyogo, Kobe 650-0047, Japan}

\author{Hiroshi Iyetomi}
\affiliation{Department of Mathematics, Niigata University, Niigata 950-2181, Japan}

\author{Hideaki Aoyama}
\affiliation{Graduate School of Sciences, Kyoto University, Kyoto 606-8502, Japan}

\author{Mauro Gallegati}
\affiliation{Dipartimento di Scienze Economiche e Sociali, Universit\`a Politecnica Delle
Marche, Piazza Martelli, 8. 60121 Ancona, Italy}

\author{Rosario N. Mantegna}
\affiliation{Center for Network Science and Department of Economics, Central European University, Nador 9, H-1051, Budapest, Hungary}
\affiliation{Dipartimento di Fisica e Chimica, Universit\`a di Palermo, Viale delle Scienze, Ed. 18, I-90128, Palermo, Italy}

\date{\today}

\begin{abstract}
We present an analysis of the credit market of Japan. The analysis is performed by investigating the bipartite network of banks and firms which is obtained by setting a link between a bank and a firm when a credit relationship is present in a given time window.  In our investigation we focus on a community detection algorithm which is identifying communities composed by both banks and firms. We show that the clusters obtained by directly working on the bipartite network carry information about the networked nature of the Japanese credit market. Our analysis is performed for each calendar year during the time period from 1980 to 2011.
Specifically, we obtain communities of banks and networks for each of the 32 investigated years, and we introduce a method to track the time evolution of these communities on a statistical basis. We then characterize communities by detecting the simultaneous over-expression of attributes of firms and banks. Specifically, we consider as attributes the economic sector and the geographical location of firms and the type of banks. 
In our 32 year long analysis we detect a persistence of the over-expression of attributes of clusters of banks and firms together with a slow dynamics of changes from some specific attributes to new ones.
Our empirical observations show that the credit market in Japan is a networked market where the type of banks, geographical location of firms and banks and economic sector of the firm play a role in shaping the credit relationships between banks and firms. 
\end{abstract}
\maketitle


\section{Introduction}
Bipartite networks are quite common in complex systems. Classic examples are networks of actors and movies, board members and companies, authors and scientific papers, etc. The customary investigation of bipartite networks is done by performing a one-mode projection for one or both of the two sets of vertices. This approach has been quite successful in the investigation of many bipartite complex systems. 
However, one-mode projection implies a certain degree of information loss that might prevent, for example, a characterization involving information about direct relationships between nodes of the two sets. 

In this paper, we investigate the bipartite network of credit relationships established between banks and firms traded at the stock exchanges and over-the-counter markets of Japan. Specifically, we aim to detect and characterize communities of banks and firms that were present in the Japanese credit market during the past years of the period of time from 1980 to 2011. Our working hypothesis is that the credit market is a networked market \cite{Easley2010}, i.e., a market where the credit relationships that are present between banks and firms are affected by attributes characterizing both banks and firms. 

Community detection in large and dense bipartite networks has been considered in the past years by several authors \cite{Guimera2007,Barber2007,Murata2009,Suzuki2009} and it is still a topic of current research \cite{Larremore2014,Melamed2014}. As for unipartite networks, community detection in bipartite networks is performed by using different approaches and different fitness measures. One widely used fitness measure is the modularity \cite{Newman2004}, i.e., the measure of the fraction of links in the network connecting vertices of the same community minus the expected value of the same quantity in the corresponding configuration model. The modularity was introduced for unipartite networks in \cite{Newman2004} and it was generalized and adapted to bipartite networks in  
\cite{Guimera2007,Barber2007,Murata2009,Suzuki2009}. The algorithms based on the generalization to the bipartite case of the modularity \cite{Guimera2007,Barber2007,Murata2009,Suzuki2009} differ among them with respect to the type of generalization. They also differ with respect to the type of communities obtained. Specifically, in Guimera et al \cite{Guimera2007} only communities with nodes of the same type are obtained. This is also the case for the algorithms of Murata \cite{Murata2009} and Suzuki and Wakita \cite{Suzuki2009} although in their case a one-to-many correspondence of each community of a specific type of nodes can be obtained.

The algorithm of Barber \cite{Barber2007} is the only one providing communities that are composed by nodes of both types and are providing a one-to-one correspondence between a group of nodes of one set and a group of nodes of the other set. In the present study, we are explicitly interested in investigating the one-to-one correspondence of groups of banks with related groups of firms. For this reason we have decided to use Barber's algorithm \cite{Note1}. 

Several complex systems can be monitored over long periods of time. The analysis and modeling of these systems can be done by considering the network connections observed for the whole time period and/or by analyzing the network in successive time intervals as, for example, daily, weekly, monthly or yearly intervals. Here we investigate the bipartite network of credit relationships between banks and firms yearly from 1980 to 2011 by investigating 32 distinct credit networks. For each year we obtain the credit network and its community structure by using Barber's BRIM (bipartite recursively induced modules) algorithm. When the time evolving nature of networks is investigated, it is important to device methods and procedures that are able to track the time evolution of specific communities of the networks also in the presence of uncertainty related to the statistical nature of the community detection process. Here we propose a method which is able to track the time evolution of communities detected in networks obtained at successive periods of time. The method uses a statistical test which is robust with respect to the heterogeneity of the size of communities and therefore works both for large and small communities.

Finally, we characterize the communities obtained for different years in terms of the over-expression of attributes of banks and firms concerning (i) the regional location of firms, (ii) economic sectors of firms, and (iii) the types of banks. The statistical validation of the over-expressed attributes is done by using a method \cite{Tumminello2011} using a multiple hypothesis test correction procedure. Our statistical validation procedure of the time evolution of communities allow us to track efficiently the evolution of the communities over time. With our approach we detect layers of networked credit relationships \cite{Easley2010} that have been present in Japan for many years. These layers of credit relationships are characterized by specific types of banks, by firms located in the same or closely related geographical regions and by firms preferentially involved in specific economic sectors.   

The paper is organized as follows. In Sect. \ref{dataset} we briefly discuss our dataset. Sect. \ref{cdbn} discusses community detection in the bipartite network of the Japanese credit market. Sect. \ref{tec} introduces a method used to track the time evolution of communities detected in networks and obtained for successive time periods. Sect. \ref{results} presents the empirical results obtained in the characterization of the over-expression of attributes of banks and firms in each community over the years and in Sect. \ref{concl} we draw our conclusions.

\section{Dataset}\label{dataset}
Our dataset is based on a survey of firms quoted in the Japanese stock-exchange markets (Tokyo, Osaka, Nagoya, in the order of
market size) and in Japanese over-the-counter (OTC) markets . The data were compiled from the firms' financial statements and survey by Nikkei Media Marketing, Inc. in Tokyo, and are commercially available \cite{Nikkei}. They include the information about each firm's borrowing obtained from financial institutions. Specifically, the dataset reports the amounts of borrowing and their classification into short-term and long-term borrowings. Long-term borrowing are considered all contracts exceeding 1 year. We examined the period 1980 to 2011, which is a time period of more than three decades. The analysis is performed yearly, and each yearly network is constructed from the dataset by using the financial statements of the considered calendar year. Since 1996 the dataset includes also OTC markets and/or JASDAQ (the present OTC market). In other studies firms of the over-the-counter market have been excluded to focus on publicly quoted firms. In the present study we investigate all firms which are present in the database.

The number of banks of the database changes year by year. It was 225 in 1980, remained approximately constant until 2001 and then decreased to 166 in 2011. The number of firms was first increasing from the value of 1414 in 1980 to the value of 3034 in 2006 and then decreasing to the value of 2706 in 2011. The number of firms increased from 1802 in 1995 to 2602 in 1996 in the presence of the first inclusion of the OTC firms in the database. During the same years the number of banks increased from 219 to 226. The density of links in the bipartite network defined as number of observed links over number of potential links was on average decreasing from the value of 0.0867 in 1980 to the vale of 0.0398 in 2011. The variation of the density of links was not too large during the first inclusion of the OTC firm. In fact the density of links decreased from 0.0721 to 0.0601 from 1995 to 1996.   

The Japanese credit market has been previously analyzed by considering one-mode projected networks \cite{DeMasi}, an eigenvalue problem determined by the weight of the credit network \cite{Fujiwara2009}, and, as in the present paper, in terms of communities detected directly on the bipartite network \cite{Iyetomi2013}. 

Concerning financial institutions, commercial banks are long-term, city, regional (primary and secondary), trust banks, insurance banks and government-related financial institutions including credit associations but excluding the Bank of Japan.  We remark that failed banks are included until the year of failure, and that merger and acquisition of banks are processed consistently to identify surviving banks. For quoted firms, those who are active in each investigated calendar year are all included even if they failed later during the considered years.

\section{Community detection in bipartite networks}\label{cdbn}
In our bipartite network a link is present between bank $i$ and firm $j$ when a credit relation (short and/or long) is present between $i$ and $j$. Links are described by a binary variable (just indicating the presence or absence or a credit relationship), i.e., in the present investigation the bipartite network is an unweighted network.

Community (cluster) detection in networks is a widely used approach used to discover empirical regularities present in a network that might be informative with respect to important aspects of the system such as its internal structure, robustness, resilience, etc. Community detection can be performed by using a series of different algorithms using different approaches and fitness measures \cite{Fortunato}.  The community detection algorithm used here is the bipartite, recursively induced modules (BRIM) algorithm, introduced in \cite{Barber2007}. It is a stochastic algorithm directly applied to the bipartite network. It uses the modularity of the bipartite network \cite{Newman2004} as a fitness measure of the partitioning procedure. 

In our analysis we have repeated the application of BRIM community detection algorithm a number of times for each year we investigate. Specifically, for each investigated year in each run we apply the algorithm 100 times and we perform 20 independent runs. 

\begin{figure}[Ht]
\begin{center}
\includegraphics[scale=0.3]{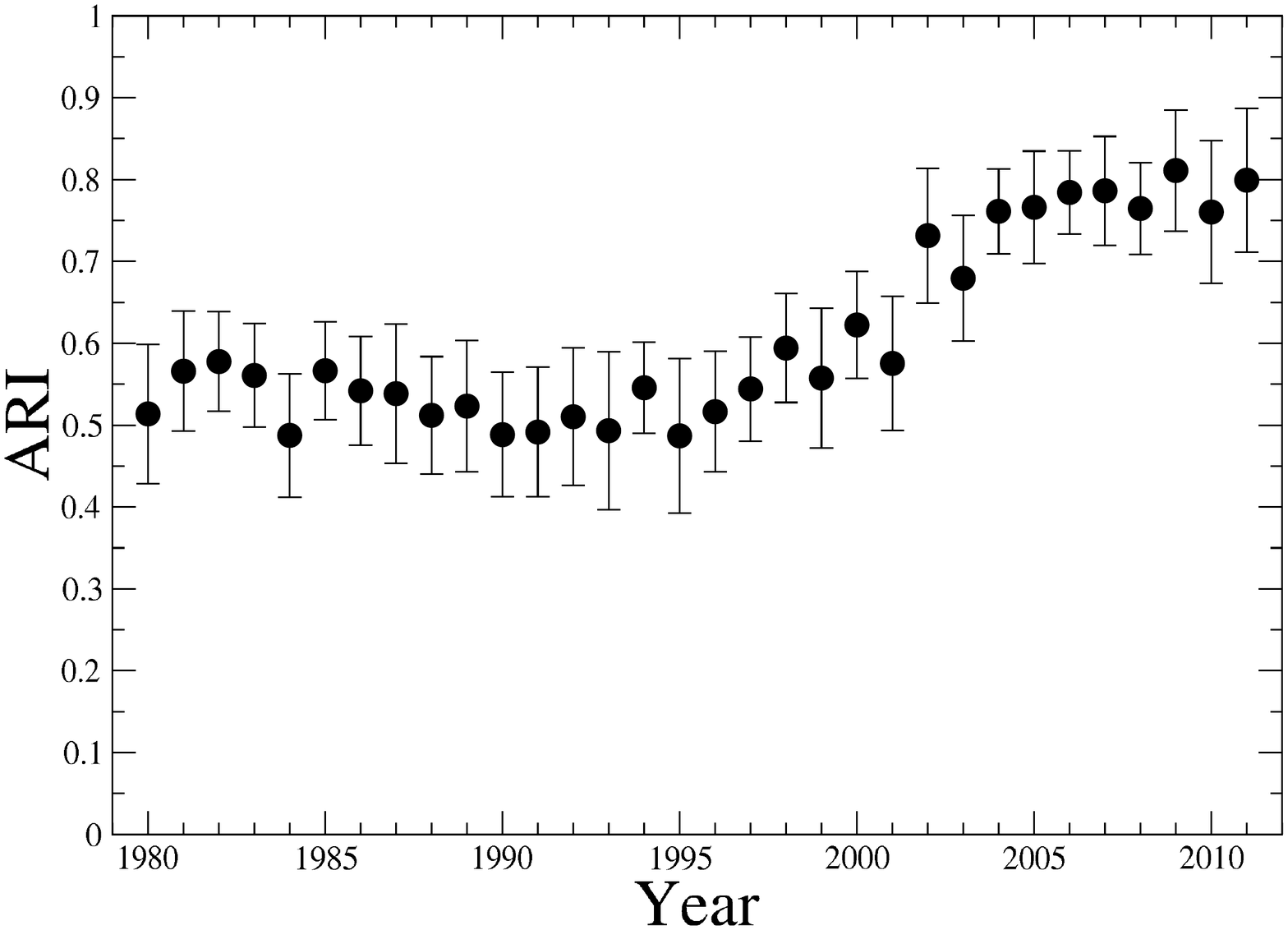}
\caption{\label{fig1} Mean value of the Adjusted Rand Index (ARI) computed between all pairs of partitions obtained in the 20 independent runs of the BRIM algorithm for each calendar year. Error bars indicate one standard deviation.}
\end{center}
\end{figure}

To quantitatively evaluate the differences which are present among the partitions obtained in the 20 independent runs performed for each calendar year, we evaluate the adjusted Rand index (ARI) \cite{Hubert1985} among all the pairs of partitions of the 20 runs. In Fig. \ref{fig1}
we show the mean value of the adjusted Rand index as a function of the calendar year. The mean value is computed for the set of 190 distinct pairs of partitions obtained from the 20 independent runs of the BRIM computed each calendar year. The error bars are one standard deviation. The adjusted Rand index is close to 0.55 from 1980 to 1995 and increases to approximately 0.8 in the time interval from 2000 to 2011. A value of the adjusted Rand index equals to one would indicate a perfect overlap of the two compared partitions whereas a value close to zero would indicate a random distribution of the nodes into the partitions. Therefore mean values ranging from 0.5 to 0.8 indicate that the different runs provide different partitions. However,  the different partitions obtained retain a significant amount of nodes within the same clusters. Moreover the degree of overlap of the partitions obtained by independent runs increases in the second half of the investigated time period. 

\begin{figure}[Ht]
\includegraphics[scale=0.3]{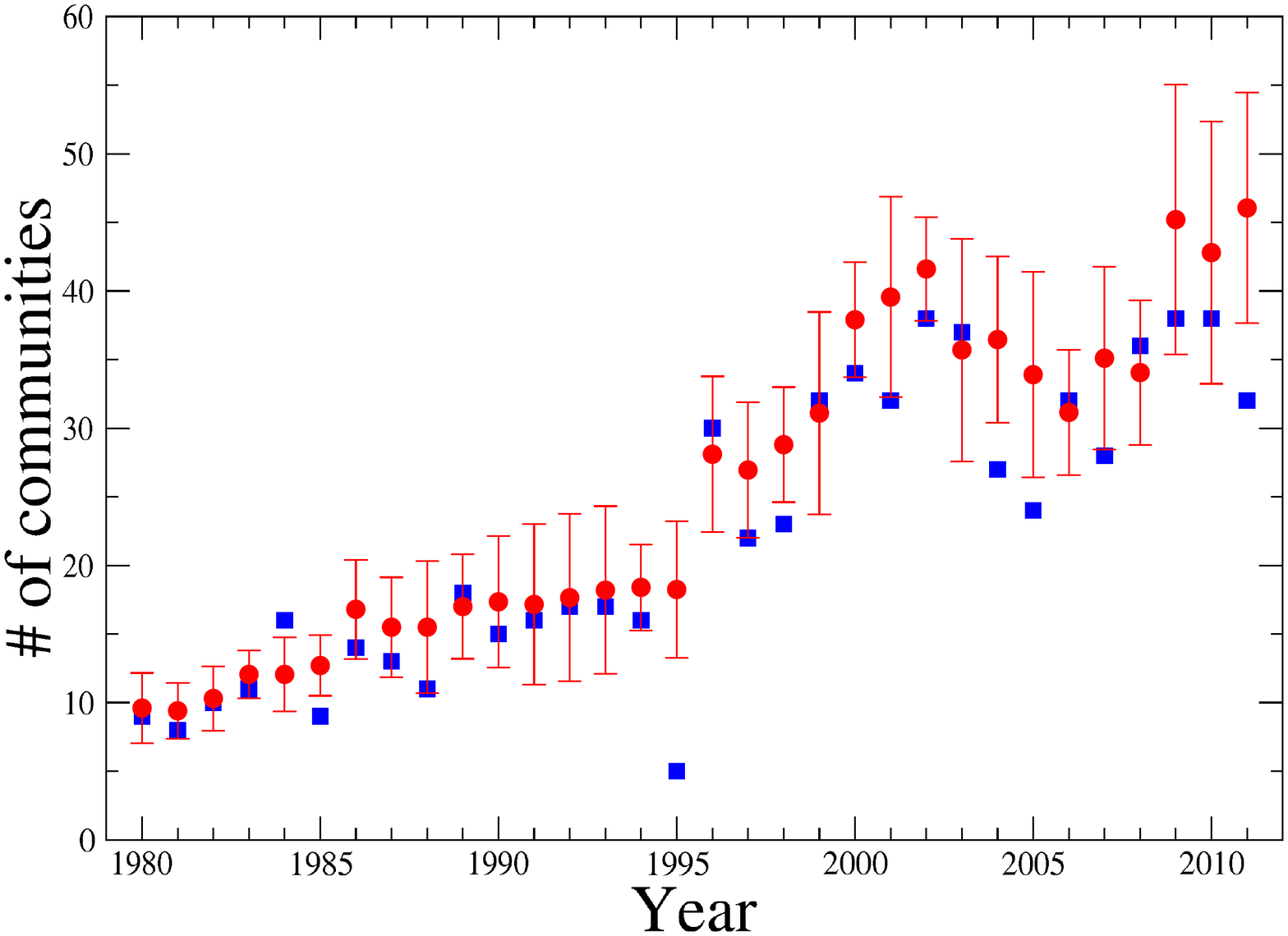}
\caption{\label{fig2} Mean value (red circle symbols) of the number of clusters obtained by applying the BRIM algorithm to the bipartite credit system bank-firm for each calendar year of the time interval 1980-2011. The mean value is obtained by considering the mean value of the number of clusters observed in the partition of best modularity obtained performing 20 different independent runs of the algorithm using random initial conditions. Error bars indicate one standard deviation. The blue symbols indicate the number of clusters obtained in the partition of the best modularity among the 20 independent runs performed.}
\end{figure}

To provide an indication of the differences observed among the partitions obtained in independent runs, in Fig. \ref{fig2} we show the time evolution of the average number of communities (red symbol) and its standard deviation obtained for each investigated year. In the figure, we also show the number of communities (blue square symbol) of the partition with the highest modularity for each year. The figure presents an overall gradual increases of the number of communities over time. The figure also shows the presence of an abrupt change of the average number of clusters that it is observed between 1995 and 1996. The reason for this abrupt change is that starting from 1996 the database is including OTC firms and therefore comprises a larger set of firms. It is worth noting that in spite of that the mean value of the adjusted Rand index (see Fig. \ref{fig1}) is not affected by the change of the size of the investigated system.

\section{Time evolution of communities}\label{tec}
The communities detected by using the BRIM algorithm and discussed in Section \ref{cdbn} are obtained year by year. It is therefore of interest to properly put communities detected on a given year in relation with communities detected in the following year. A time evolution of the communities can be detected by considering what are the communities of year $t+1$ in which one detects an over-expressed amount of elements of a given community of year $t$. The community detection procedure has a certain degree of stochasticity  and degeneracy with respect to small differences of the fitness measure and therefore the membership of an element into a certain community might also just be due to chance. We therefore need a method detecting over-expression of the same composition in communities of two successive years that is based on a carefully devised statistical procedure which is robust to the size heterogeneity of the different communities. 

Hereafter, we propose such a method. Suppose that in period $t$ there are $N_t$ communities $C_i^t$, $i=1, \cdots, N_t$ and in period $t+1$ there are $N_{t+1}$ other communities $C_j^{t+1}$, $j=1, \cdots, N_{t+1}$. For all the $C_i^t$ communities of period  $t$  we search amongst all $N_{t+1}$ communities of period $t+1$ which communities $C_{j}^{t+1}$ have an over-represented composition of elements also present in a community at time $t$. Specifically, let us call $n_i^{t}$ the number of elements in $C_i^t$, $n_j^{t+1}$ the number of elements in $C_j^{t+1}$ and $n_{ij}^{t,t+1}$ the number of common element between $C_i^t$ and $C_j^{t+1}$. Let us call $N^{t,t+1}$ the number of distinct vertices in the two consecutive periods $t$ and $t+1$. The probability that $n_{ij}^{t,t+1}$ is observed by chance is given by the hypergeometric distribution $H(n_{ij}^{t,t+1} | N^{t,t+1}, n_i^{t}, n_j^{t+1})$ where:
\begin{eqnarray}
                                H(X|N,M,K)=\frac{{M \choose X} {N-M \choose K-X}}{{N \choose K}}. \label{hyp}
\end{eqnarray}
Therefore for each pair of clusters we can compute a $p$-value
\begin{eqnarray}
                                p_{ij}^{t,t+1}=1- \sum_{x=0}^{n_{ij}^{t,t+1}-1} H(x | N^{t,t+1}, n_i^{t}, n_j^{t+1}). \label{pval}
\end{eqnarray}
After setting the appropriate $p$-value threshold $p_t$, the above methodology gives us a way to select the communities in year $t+1$ that are linked to a given community in year $t$ in a statistically robust way. 

To avoid the presence of false positive, the $p$-value threshold must be corrected to take into account that we are performing a multiple hypothesis test comparison. Indeed, for each pair of consecutive periods we perform the test $N_t \cdot N_{t+1}$ times against the null hypothesis of random distribution of elements among two partitions of communities of consecutive periods. 
Moreover we perform these tests for all pairs of consecutive years in our dataset, i.e., from $1980$ to $2011$. The most restrictive multiple hypothesis test correction is the Bonferroni correction, which prescribes that the modified $p$-value threshold $p_B$ is:
\begin{eqnarray}
                          p_B=p_t / \bigl(\sum_{t=1980}^{2011-1} N_t \cdot N_{t+1} \bigl).
\end{eqnarray}
In the present investigation we have set  $p_t=0.01$. 

\begin{figure}
\begin{center}
\includegraphics[scale=0.22]{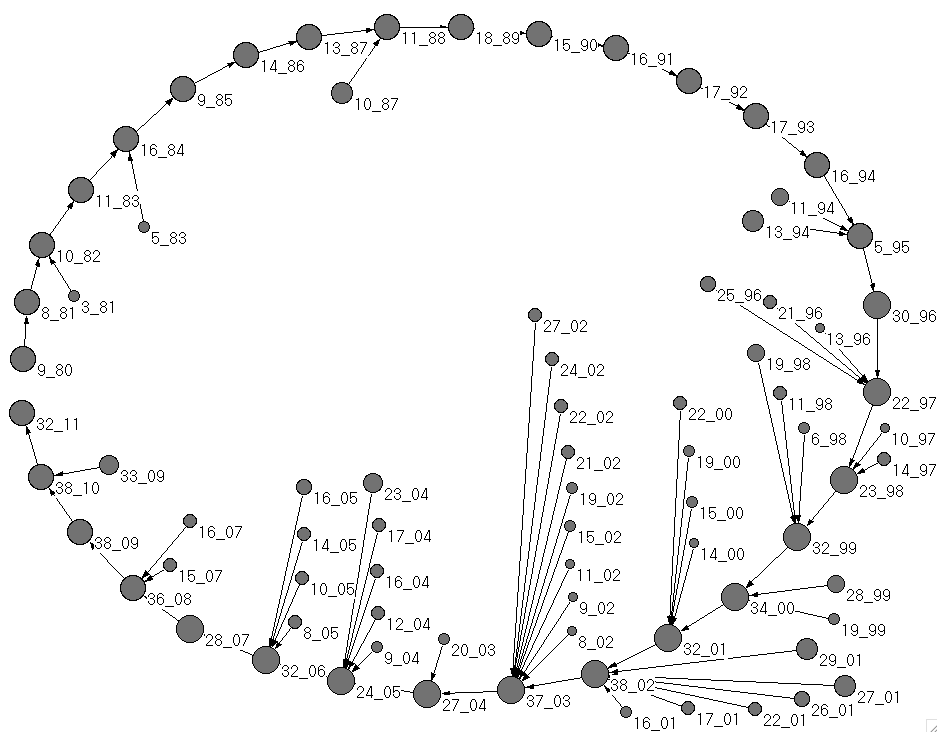}
\caption{\label{figE} Graphical representation of the interrelationship of clusters detected in successive years. The figure shows only the statistical validations observed starting from the largest community of 1980 (labeled as 9\_80) and considering validation between all pairs of clusters observed for each pair of successive years. The size of the vertex symbol is proportional to the logarithm of the size of the cluster.  The figure shows that the statistical validation of the cluster composition clearly show that the largest cluster always evolves into the largest cluster of the successive year.}
\end{center}
\end{figure}

In Fig. \ref{figE} we show a graphical representation of the interrelationships of communities that are statistically validated in successive years. The graphical representation is the time evolution of the biggest community of 1980 (labeled as 9\_80). The size of each vertex is proportional to the logarithm of the size of the community.  The statistical validation procedure shows that the largest community of year $t$ evolves into the largest community of year $t+1$ for all the investigated years. In addition to this primary channel of community evolution we also detect that in some years other smaller communities merge part of them into the largest one (this process is more pronounced during the years 2000, 2001 and 2002). For the sake of clarity, among the communities merging into the largest community, only communities at one year distance from the largest community of each year are shown in the figure. In the following section we will investigate the over-expression of the attributes characterizing the elements of the largest community observed in each calendar year.

\begin{figure}
\begin{center}
\includegraphics[scale=0.2]{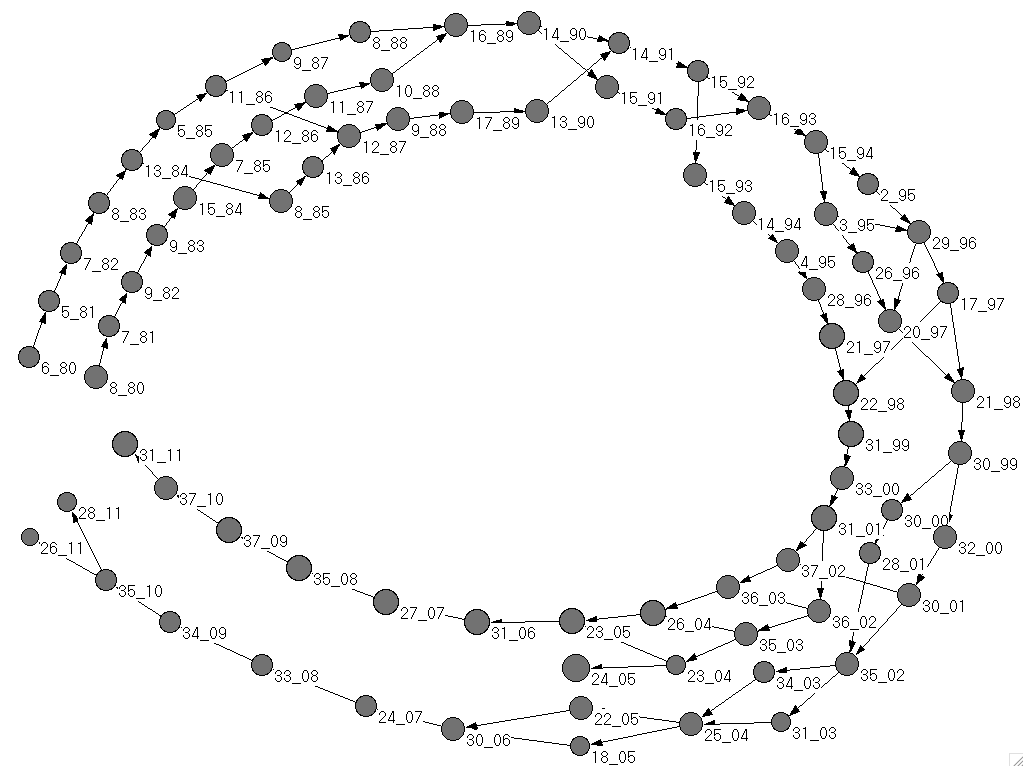}
\caption{\label{figD} Graphical representation of the interrelationship of clusters detected in successive years. The figure shows only the statistical validations observed starting from the second and the third largest communities of 1980 (labeled as 6\_80 and 8\_80) and considering validation between all pairs of clusters observed for each pair of successive years.
Only statistically validated directed connection from each cluster to the ones of the successive year are shown. The incoming validated connection from other clusters of the previous year are not shown to make the figure readable.   The size of the vertex symbol is proportional to the logarithm of the size of the cluster.  The figure shows that the statistical validation of the clusters presents the evolution of three main branches. One of these branches merges into the evolution of the largest in 2005 (see the evolution of 23\_04 in 24\_05).}
\end{center}
\end{figure}

In Fig. \ref{figD} we track the evolution of the second and the third largest communities of 1980 (labeled as 6\_80 and 8\_80). In this case the evolution of these communities presents three main branches shown in the figure as parallel evolving branches. However, splitting and coalescence of the branches are observed over time. In this figure we show only ``forward" community evolution, i.e., we show all the validated relationships between communities shown at year $t$ with communities at year $t+1$ but, differently than in Fig. \ref{figE} we do not show validated relationships between communities at year $t+1$ and communities at year $t$ different from the one already shown in the figure.
The additional incoming validated connections from other communities of the previous year are not shown to make the figure readable.  As in Fig. \ref{figE}, the size of each vertex symbol is proportional to the logarithm of the size of the community.  The over-expression of attributes of elements belonging to the communities of the three main branches will be discussed in the following Section.

\section{Over-expression of attributes}\label{results}
The identification of bank-firm partitions and the statistical validation of their time evolution provides the basis for the understanding of the networked structure of the credit system and its time evolution. A further step is to look for information characterizing the obtained clusters and their time evolution. In other words it is important to characterize the clusters in terms of attributes over-expressed by the elements belonging to the same clusters with respect to an appropriate random null hypothesis. The method used is illustrated in \cite{Tumminello2011}. It should be noted that the null hypothesis takes into account the heterogeneity of the tested attributes and therefore the over-expressed attributes are not necessarily the most common ones in each cluster, but rather are those whose frequency in the cluster is over-expressed with respect to a null hypothesis taking into account the heterogeneity of attributes. In our analysis we account for multiple hypothesis test correction by using the Bonferroni correction.

The metadata available for the characterization of firms and banks allows us to identify the economic sector and the prefecture of the main office of firms and the type of bank. In Table \ref{tab1} we 
summarize the over-expressed attributes observed for the time evolution of the largest community (see Fig. \ref{figE}).  In the Table  we provide the calendar year, the number of banks in the cluster, the number of firms in the cluster, and the over-expressed (i) prefectures where firms are located, (ii) economic sectors of the firms and (iii) types of banks. We notice that the type of bank over-expressed in this cluster is the type labeled as ``City banks" for the majority of the investigated years. These banks are large commercial banks operating in the entire country. The fact that the over-expression of ``City banks" is not observed after 2005 does not mean that the role of City banks is no more present in those years. In fact also for those years we detect a significant number of City banks in the considered cluster. The reason why this bank category start to be not over-expressed lays in the fact that the number of ``City banks" is declining over time (due to merging) and the validation procedure is conducted at the most severe level of multiple hypothesis test correction. In fact the Bonferroni threshold used to validate the over-expression is set to $0.01/R_t$ where 0.01 is the univariate threshold and $R_t=(N_S+N_P+N_B) \cdot N_t$ is the total number of tests done in the statistical validation of communities of the year $t$.  More specifically, $N_S$ is the number of distinct economic sectors, $N_P$ is the number of distinct Japanese prefectures, $N_B$ is the number of types of banks, and $N_t$ is the number of communities detected at year $t$. In this way, we minimize the number of false positive but unavoidably increase the number of false negative. 

The over-expressed prefectures are the prefectures of Kanagawa (14) and Tokyo (13), i.e. two prefectures of the so-called greater Tokyo area.  The Table also shows the over-expression of the main economic sectors of the firms belonging to the community. The over-expressed economic sectors are Electric and electronic equipment (EEE) for the time period 1980-1993, and Services (S) and Wholesale trade (WT) for the time period 1996-2011.

\begin{widetext}

\begin{table}
\caption{\label{tab1} Summary of information about the largest cluster detected by the BRIM algorithm in each calendar year. In the table for each cluster we report the year, the number of banks, the number of firms, the over-expressed Japanese prefecture of firms (the information is provided in terms of the standard 2 digit code), the over-expressed economic sector, and the over-expressed bank type. According to the 2 digit prefecture code we have: 13 Tokyo and 14 Kanagawa. The over-expressed economic sectors are Electric and Electonic Equipments (EEE), Services (S), and Wholesale trade (WT). The over-expressed type of bank is ``city banks" (CB).}
\begin{center}
\begin{tabular}{l|rr|ccc|r|rr|ccc}
  \hline
  Year & banks & firms & prefecture	& sector & bank type	& Year & banks & firms & prefecture	& sector & bank type\\
  \hline
 1980 & 23 & 557 & 14 & EEE & CB & 1996 & 18 & 975 & 13~~14 & S & CB\\
 1981 & 18 & 514 & 14 & EEE & CB & 1997 & 14 & 971 & 13~~14 & S & CB\\ 
 1982 & 21 & 534 & 14 & --      & CB &  1998 & 16 & 1069 &  13~~14 & S & CB\\
 1983 & 21 & 560 & 14 & --      & CB &  1999 & 14 & 1104 & 13~~14 & S, WT & CB\\
 1984 & 16 & 561 & 14 & EEE & CB &  2000 & 11 & 959 & 13~~14 & S, WT & CB\\
 1985 & 20 & 561 & 14 & EEE & CB &  2001 & 9 & 865 & 13 & S, WT & CB\\
 1986 & 18 & 564 & -- & -- & CB &  2002 & 9 & 917 & 13 & S, WT & CB\\
 1987 & 16 & 555 & -- & -- & CB &  2003 & 8 & 891 & 13 & S, WT & CB\\
 1988 & 20 & 611 & -- & EEE & CB &  2004 & 11 & 912 & 13 & S, WT & --\\
 1989 & 19 & 613 & 14 & EEE & CB &  2005 & 6 & 905 & 13 & S, WT & CB\\
 1990 & 21 & 643 & 13 & -- & CB &  2006 & 8 & 902 & 13~~14 & S, WT & --\\
 1991 & 20 & 664 & -- & S & CB &  2007 & 6 & 857 & 13~~14 & S, WT & --\\
 1992 & 19 & 614 & 13~~14 & -- & CB &  2008 & 7 & 813 & 13~~14 & S, WT & --\\
 1993 & 15 & 638 & 13 & EEE & CB &  2009 & 11 & 787 & 13~~14 & S, WT & --\\
 1994 & 18 & 670 & 13 & -- & CB & 2010 & 11 & 748 & 13~~14 & S, WT & --\\
 1995 & 20 & 691 & -- & -- & CB &  2011 & 8 & 725 & 13~~14 & S, WT & --\\
 \hline
\hline
\end{tabular}
\end{center}
\end{table}

\end{widetext}

Table \ref{tab1} shows two pronounced changes in the number of firms belonging to the main clusters. The main change (also observable in term of average number of clusters detected by the BRIM algorithm in Fig. \ref{fig2}) occurs in 1996 which is the first year of inclusion of firms traded in the OTC markets in the database. The second change is observed for the period 1999-2001. In fact, starting from 2000 the database reports credit information covering a fraction of the credit close to approximately 20-30\% of the total credit being referred to as ``unknown - other financial institutions". This form of credit involves approximately 300 firms both publicly quoted and traded in the OTC markets. In other words starting from 2000 a large number of firms of the database receive their credit from ``unknown - other financial institutions". This set of firms makes a rather stable star-like cluster.
Such a community is detected from the BRIM algorithm systematically since its first formation in 2000. Most probably the presence of this community and its stability is the main source of the increased mean value of the adjusted Rand index observed after 2000 in Fig. \ref{fig1}.   

In Fig. \ref{figD} we have shown the time evolution of the second and third largest communities of 1980. In this case we observe an evolution of the communities that on average presents  three main branches characterized by the over-expression of several Japanese prefectures, economic sectors and type of banks. All over-expression are summarized in Table \ref{STC} where we note three main branches of community. The first one starts in 1980 and last until 2011 (see the first column of Table \ref{STC}). It presents over-expression of firms of economic sectors Utilities (U) and Credit Leasing (L). The over-expressed banks are Life-insurance banks (LI) and Insurance banks (IB) banks. The over-expression of Utilities is observed until 2000. Starting from 2000 only firms belonging to the Credit Leasing economic sector are over-expressed. For this branch of clusters the geographical location of firms shows that the Japanese prefectures of Tokyo (labeled as 13), Hiroshima (34) and Fukuoka (40) are over-expressed in several years. During the most recent years several prefectures of the southern part of Japan (e.g. prefectures labeled as 28 (Hyogo), 33 (Okayama), 34, 37 (Kagawa) and 40) are over-expressed.

The second branch starts in 1980 and ends approximately in 2005  (see the second column of Table \ref{STC}). This second branch presents over-expression of firms of the Construction (C) economic sector and of the Regional banks (RB) and occasionally of the Second regional banks (SR).  The geographical over-expression points out Japanese prefectures of Hiroshima and Fukuoka and of Tokyo in a few cases. 

The third branch starts in 1985 and ends in 2011  (see the third column of Table \ref{STC}). In this last case the branch presents persistent over-expression of firms of the Railroad Transportation (RT) and Chemicals (Ch) sectors. An over-expression of banks classified as Life-insurance banks (LI) is observed after 1997.  The geographical over-expression mainly involves the prefecture of Tokyo especially during the most recent years.

In summary we observe three distinct branches well characterized over time by a rather stable over-expressions of economic sector and type of banks. Also the over-expression of the regional location of firms, although less stable than the ones of the economic sector and of the type of bank, shows a high degree of persistence over time. The clusters of banks and firms detected by the BRIM are able to detect a networked nature of the Japanese credit market with a time scale of the dynamics of the communities covering several years.
 
\begin{widetext}

\begin{table}
\caption{\label{STC} Summary of information about the evolution of a few large clusters detected by the BRIM algorithm in each calendar year. The evolution follows the scheme shown in Fig. \ref{figD}. In the table for each community we report the code of the community (id\_year), the number of banks, the number of firms, the over-expressed Japanese prefecture of firms (the information is provided in terms of the standard 2 digit code), the over-expressed economic sector of firms, and the over-expressed bank type. The 2 digit prefecture code is the one of Japan's International Organization for Standardization, and it can be found online at the web page Prefectures of Japan in Wikipedia.  
The over-expressed economic sectors are Construction (C), Credit Leasing (CL), Chemicals (Ch), Electric and Electronic Equipments (EEE), Motor parts (MV),  Railroad Transportation (RT), Sea Transportation (ST), Services (S), Utitilies (U) and Wholesale trade (WT). The over-expressed type of bank are ``city banks" (CB), Life-insurance banks (LI), Regional banks (RB), Insurance banks (IB), and Second regional banks (SR).}
\begin{center}
\scalebox{0.8}{
\begin{tabular}{l|rr|ccc|r|rr|ccc|r|rr|ccc}
  \hline
  cluster & $N_b$ & $N_f$ & pref.	& sector & B.t.	&  cluster & $N_b$ & $N_f$ & pref.	& sector & B.t.	&  cluster & $N_b$ & $N_f$ & pref.	& sector & B.t.	\\
  \hline
 6\_80 & 45 & 241 & -- & -- & LI~~IB & 8\_80 & 128 & 305 & -- & -- & RB~~SR & ~ & ~ & ~ & ~ & ~ & ~ \\
 5\_81 & 47 & 201 & -- & RT~~U & LI~~IB & 7\_81 & 97 & 214 & 13 & C & RB & ~ & ~ & ~ & ~ & ~ & ~ \\ 
 7\_82 & 49 & 194 & -- & U  & LI~~IB &  9\_82 & 89 & 208 &  13 & C & RB & ~ & ~ & ~ & ~ & ~ & ~ \\
 8\_83 & 46 & 166 & 13 & U  & LI~~IB &  9\_93 & 85 & 211 & -- & C & RB & ~ & ~ & ~ & ~ & ~ & ~ \\
 13\_84 & 47 & 146 & -- & U & LI~~IB &  15\_84 & 111 & 277 & -- & -- & RB & ~ & ~ & ~ & ~ & ~ & ~ \\
 5\_85 & 41 & 102 & -- & -- & IB &  7\_85 & 85 & 260 & -- & -- & RB & 8\_85 & 29 & 381 & -- & RT & -- \\
11\_86 & 48 & 164 & -- & U & LI~~IB & 12\_86 & 80 & 228 & -- & C & RB & 13\_86 & 21 & 292 & -- & -- & -- \\
 9\_87 & 52 & 108 & 13 & CL~~U & LI~~IB &  11\_87 & 89 & 235 & 40 & C & RB & 12\_87 & 16 & 319 & -- & RT & -- \\
  8\_88 & 58 & 203 & 13 & U & LI~~IB &  10\_88 & 113 & 314 & 40 & C & RB~~SR & 9\_88 & 21 & 339 & -- & RT & ~ \\
 16\_89 & 121 & 225 & 34~~40 & CL~~U & IB & ~ & ~ & ~ & ~ & ~ & ~ & 17\_89 & 21 & 367 & -- & RT & -- \\
 14\_90 & 135 & 220 & 13 & CL & -- &  ~ & ~ & ~ & ~ & ~ & ~ & 13\_90 & 15 & 336 & -- & RT & -- \\
 14\_91 & 51 & 237 & -- & CL~~U & LI~~IB &  15\_91 & 93 & 223 & 34~~40 & C & RB & ~ & ~ & ~ & ~ & ~ & ~ \\
 15\_92 & 52 & 188 & 13 & CL & LI~~IB &  16\_92 & 78 & 199 & 34~~40 & C & --& ~ & ~ & ~ & ~ & ~ & ~ \\
 16\_93 & 135 & 273 & 13~~40 & CL & ~ &  ~ & ~ & ~ & ~ & ~ & ~ & 15\_93 & 11 & 309 & -- & RT & -- \\
 15\_94 & 126 & 246 & 13 & CL~~U & IB & ~ & ~ & ~ & ~ & ~ & ~ & 14\_94 & 13 & 322 & -- & RT & -- \\
 2\_95 & 57 & 169 & -- & CL~~U & LI~~IB &  3\_95 & 102 & 325 & 34~~40 & C & RB & 4\_95 & 20 & 486 & -- &Ch~~RT~~ST & --  \\
29\_96 & 45 & 241 & -- & -- & LI~~IB & 26\_96 & 128 & 305 & -- & -- & RB~~SR & 28\_96 & ~ & ~ & ~ & ~ & ~ \\
17\_97 & 71 & 149 & 13 & CL~~U & IB & 20\_97 & 72 & 361 & 1~~15~~34~~40 & C & RB & 21\_97 & 17 & 524 & -- & Ch~~MV~~RT 
& -- \\
21\_98 & 119 & 342 & 34~~40 & C~~CL & -- & ~ & ~ & ~ & ~ & ~ & ~ &  22\_98 & 34 & 499 & -- & Ch~~RT & LI \\
30\_99 & 119 & 377 & 34~~40 & -C~~CL~~U & -- & ~ & ~ & ~ & ~ & ~ & ~ &  31\_99 & 24 & 550 & 26~~27 & Ch & -- \\
32\_00 & 92 & 345 & 10~~13 & CL & IB & 30\_00 & 42 & 226 & 33~34~38~40~43~46 & -- & -- &  33\_00 & 19 & 477 & 26~~27 & Ch & ~ \\
30\_01 & 93 & 232 & 7 & C~~CL & IB & 28\_01 & 34 & 228 & 33~34~35~37~38~40 & -- & -- &  31\_01 & 16 & 574 & 14 & Ch & -- \\
35\_02 & 82 & 278 & 34~37~40 & -- & -- &  36\_02 & 11 & 379 & -- & -- & -- & 37\_02 & 26 & 383 & -- & -- & -- \\
\hline
34\_03 & 69 & 241 & 27~28~33~34~37 & CL & -- & 35\_03 & 8 & 326 & 13 & -- & -- & 36\_03 & 26 & 450 & 22 & -- & LI \\
31\_03 & 21 & 110 & 40~46 & -- & -- & ~ & ~ & ~ & ~ & ~ & ~ & ~ & ~ & ~ & ~ & ~ & ~ \\
\hline
25\_04 & 85 & 338 & 33~34~37~40 & CL & -- & 23\_04 & 6 & 151 & 14 & -- & -- & 26\_04 & 28 & 550 & 13 & Ch & -- \\
\hline
22\_05 & 84 & 310 & 1~28~33~34~37 & CL & -- & 24\_05 & 6 & 905 & 13 & S~~WT & CB & 23\_05 & 26 & 602 & 15 & RT & LI \\
18\_05 & 18 & 109 & 40~46 & -- & -- &  ~ & ~ & ~ & ~ & ~ & ~ & ~ & ~ & ~ & ~ & ~ & ~ \\
\hline
30\_06 & 71 & 306 & 28~33~34~37~40 & CL & -- & ~ & ~ & ~ & ~ & ~ & ~ & 31\_06 & 23 & 646 & 13 & RT & LI \\
24\_07 & 27 & 174 & 34~35~40~46 & -- & -- & ~ & ~ & ~ & ~ & ~ & ~ & 27\_07 & 24 & 662 & 13 & Ch~~RT & -- \\
33\_08 & 51 & 262 & 28~33~34~37~40 & -- & -- & ~ & ~ & ~ & ~ & ~ & ~ & 35\_08 & 30 & 595 & 13 & CL~~RT & -- \\
34\_09 & 34 & 225 & 28~33~34~37~40 & -- & -- & ~ & ~ & ~ & ~ & ~ & ~ & 37\_09 & 29 & 546 & 13 & -- & LI \\
35\_10 & 42 & 210 & 28~33~34~37~40 & -- & -- & ~ & ~ & ~ & ~ & ~ & ~ & 37\_10 & 22 & 483 & 13 & RT & LI \\
\hline
28\_11 & 17 & 154 & 28~33~34~37 & -- & -- & ~ & ~ & ~ & ~ & ~ & ~ & 31\_11 & 23 & 514 & 13 & RT & -- \\
26\_11 & 18 & 89 & 40~46 & -- & -- & ~ & ~ & ~ & ~ & ~ & ~ & ~ & ~ & ~ & ~ & ~ & ~ \\
\hline 
\hline
\end{tabular}
}
\end{center}
\end{table}

\end{widetext}

\section{Conclusions}\label{concl}
In our study we analyze the time evolution of the bank-firm credit relationships in Japan over a period of time longer than 30 years. The analysis is performed on the bank-firm credit network observed yearly. The bipartite network is analyzed yearly and the communities of banks and firms are characterized with respect to the over-expression of firms' economic sectors, firms' Japanese prefectures and types of banks. 

In our study it was crucial to select a community detection algorithm directly working on the bipartite network that is providing communities composed by both types of vertices (banks and firms). The choice of a one-to-one correspondence between banks' partitions and firms' partitions also simplify our analysis of simultaneous over-expression of attributes of both banks and firms. With this approach we have been able to show the existence of layers of the credit market involving groups of firms characterized by specific economic sectors and regional locations (prefectures) and specific types of banks. These empirical observations show that the credit market in Japan is a networked market.

The robustness of our results is shown by the ability of our approach in detecting both the long term stability and the slow dynamics of the detected communities. The time evolving communities have been tracked from each year to the next one by using a newly introduced statistical method able to track the time evolution of communities detected in successive periods of time also in the presence of size heterogeneity of the communities. It is worth noting that our method presently used to track time evolution of communities can also be easily adapted to link communities detected in a multiplex network. 


{\bf Acknowledgments} LM, SM, MG and RNM acknowledge support from the EU project CRISIS and the INET project ``New tools in the credit network modeling with agents' heterogeneity". This work is partially supported by Grant-in-Aid for Scientific Research (KAKENHI) Grant Numbers 24243027, 25282094, 25400393 by JSPS, the Program for Promoting Methodological Innovation in Humanities and Social Sciences by Cross-Disciplinary Fusing of the JSPS, and by the European Community Seventh Framework Programme (FP7/2007-2013) under Socio-economic Sciences and Humanities, grant agreement no. 255987 (FOC-II) and 297149 (FOC-INCO). LM and RNM wish to thank D. Larremore for discussion and for sharing a program of community detection based on a bipartite stochastic block model.

\end{document}